\documentclass[twocolumn,superscriptaddress,showpacs,preprintnumbers,amsmath,amssymb]{revtex4}

\usepackage{latexsym}% Basic package(s)
\usepackage{graphicx}% Include figure files
\usepackage{dcolumn}% Align table columns on decimal point

\newcommand{\bra}[1]{\langle #1 |}
\newcommand{\keta}[1]{| \hspace{-0.1cm} #1 \rangle}
\newcommand{\ket}[1]{| #1 \rangle}

\newcommand{\braket}[2]{\langle #1 | #2 \rangle}

\def\etal{\textit{et al.{}}}
\newcommand{\ua}{\uparrow}
\newcommand{\da}{\downarrow}
\newcommand{\im}{\dot{\iota}\,}
\begin{document}

\title{Generation of four-partite GHZ and
       W states by using a high-finesse bimodal cavity}

\author{D. Gon\c{t}a}
\email{gonta@physi.uni-heidelberg.de}
\affiliation{Max--Planck--Institut f\"{u}r Kernphysik, Postfach
             103980, D--69029 Heidelberg, Germany}

\author{S. Fritzsche}
\email{s.fritzsche@gsi.de}
\affiliation{Max--Planck--Institut f\"{u}r Kernphysik, Postfach
             103980, D--69029 Heidelberg, Germany}
\affiliation{Gesellschaft f{\"u}r Schwerionenforschung, D-64291 Darmstadt, Germany}%

\author{T. Radtke}
\email{tradtke@physik.uni-kassel.de}
\affiliation{Institut f\"{u}r Physik, Universit\"{a}t Kassel,
             Heinrich-Plett-Str. 40, D-34132 Kassel, Germany}

\date{\today}

\begin{abstract}
We propose two novel schemes to engineer four-partite entangled
Greenberger-Horne-Zeilinger (GHZ) and W states in a deterministic
way by using chains of (two-level) Rydberg atoms within the framework
of cavity QED. These schemes are based on the resonant interaction
of the atoms with a bimodal cavity that simultaneously supports,
in contrast to a single-mode cavity, two independent modes of the
photon field. In addition, we suggest the schemes to reveal the
non-classical correlations for the engineered GHZ and W states. It
is shown how these schemes can be extended in order to produce
general N-partite entangled GHZ and W states.
\end{abstract}

\pacs{42.50.Pq, 42.50.Dv, 03.67.Mn}

\maketitle

\section{Introduction}

Entanglement is known today as a key feature of quantum mechanics;
it has been found important not only for studying the nonlocal and
nonclassical behavior of quantum particles but also for applications
in quantum engineering and quantum information theory \cite{nc},
such as super-dense coding \cite{bew}, quantum cryptography
\cite{eke}, or for the search of quantum algorithms \cite{gro}.
Apart from the Bell states as prototype of two-partite entangled
states, several attempts have been made recently in order to create
and control `entanglement' also for more complex states. Owing to
the fragile nature of most of these states, however, their
manipulation still remains a challenge for experiment and only a few
`proof-of-principle' implementations have been realized so far that
generate entangled states with more than two parties in a
well-defined way. For systems of three parties, for example, one
often considers the Greenberger-Horne-Zeilinger (GHZ) state
\cite{ghz}
\begin{eqnarray}\label{3-ghz}
\ket{\Psi_{\rm\, GHZ}^{(3)}} & = &
\frac{1}{\sqrt{2}} \left( \ket{\ua_1, \ua_2, \ua_3} +
                          \ket{\da_1 ,\da_2 ,\da_3} \right)
\end{eqnarray}
and the W state \cite{pra62}
\begin{eqnarray}\label{3-w}
\ket{\Phi_{\rm\, W}^{(3)}} & = &
\frac{1}{\sqrt{3}} \left( \ket{\ua_1,\da_2, \da_3} +
                          \ket{\da_1 ,\ua_2 ,\da_3} + \right.
\nonumber  \\ & &
\hspace{1.2cm} \left. + \ket{\da_1 ,\da_2 ,\ua_3} \right)
\end{eqnarray}
as two kinds of pure entangled three-qubit states. In this notation,
as usual, $\keta{\ua_n}$ and $\keta{\da_n}$ refer to the two
distinguishable `projections' of qubit $n$, such as its spin,
excitation state, polarization or some other `two-level' property of
a given quantum system, while $\ket{\ua_1, \ldots, \ua_n}$ denotes
the direct pro\-duct of states $\ket{\ua_1}, \ldots, \ket{\ua_n}$,
respectively. The GHZ and W states are known also as
\mbox{\textit{genuine}} entangled states since they cannot be
transformed into each other under \textit{local operations and
classical communication} (LOCC) protocols \cite{pra62}. Indeed, the
properties of these states have been explored in details during
recent years with regard to different quantum measures, separability
criteria, or concerning the violation of local realism \cite{ghz,
prl65, prl67}. For instance, while the GHZ state is fragile under
qubit loss, leading to a separable quantum state if just one of the
three qubits is traced out, the three-partite W state still results
in the Bell state $\frac{1}{\sqrt{2}} \left( \ket{\ua_1,\da_2} +
\ket{\da_1 ,\ua_2} \right)$ when the third qubit is projected upon
the state $\keta{\da_3}$. Various experiments have been reported in
the literature for generating three-qubit GHZ and W states by
applying optical systems \cite{prl_nat, prl_jmo}, \mbox{nuclear}
magnetic resonance \cite{pra61, pra66}, cavity QED \cite{sc288, no},
or ion trapping techniques \cite{sc304}.

In the framework of cavity QED in particular, in which neutral atoms
couple to a high-finesse microwave cavity, Rauschenbeutel and
coworkers \cite{sc288} prepared the excited states of three
two-level Rydberg atoms (using circular atomic states which
correspond to levels $n$ and $n+1$) in an entangled GHZ state
(\ref{3-ghz}) by utilizing a single-mode superconducting cavity
\cite{haroche, haroche1}. In these experiments, the cavity field
mediates the interactions between the atoms that pass successively
through the cavity, and the control over the light fields and atoms
(atomic chain) is achieved owing to the high quality of the cavity.
In the language of these cavity experiments, usually two parts of
the measurements are distinguished: The (so-called)
\textit{longitudinal} experiment to prepare the entangled state of
the Rydberg atoms, and the \textit{transversal} experiment that
helps to \mbox{`reveal'} the produced entanglement. This latter part
is realized by observing the `non-classical' correlations for a
series of projective measurements on the population of the Rydberg
states after the given chain of atoms has passed through the cavity
and the atomic \mbox{Ramsey} interferometer (see Section III). The
experiments by Rauschenbeutel \etal{} \cite{sc288} nicely
demonstrated the possibility of using an atom-cavity (quantum) phase
gate in order to entangle three atomic qubits, and it has triggered
the community to undertake steps towards the controlled manipulation
of multi-partite entangled states. Up to the present, however, only
a few case studies are known \cite{prl86, nat404, prl90}, where the
four-partite entangled states have been generated by using optical
systems and ion trapping techniques.

Of course, a detailed analysis is required for every particular
realization of $N-$partite quantum systems in order to  work out an
\textit{experimental scheme} that enables one to generate and
observe reliably entanglement in the system. From the viewpoint of
theory, such a scheme can be understood also as a quantum circuit
or, simply, a (temporal) sequence of steps for dealing with the
individual parts (qubits) of the system. In the present work, we
suggest two (experimentally feasible) schemes for generating the
four-partite GHZ and W states
\begin{eqnarray}
\label{4-ghz}
\ket{\Psi_{\rm\, GHZ}^{(4)}} & = &
\frac{1}{\sqrt{2}} \left( \ket{\ua_1,\ua_2 ,\ua_3, \ua_4} +
                          \ket{\da_1 ,\da_2 ,\da_3 ,\da_4} \right),
\\[0.1cm]
\ket{\Phi_{\rm\, W}^{(4)}} & = &
\frac{1}{2} \left( \ket{\ua_1, \da_2 ,\da_3 ,\da_4} +
                   \ket{\da_1 ,\ua_2, \da_3, \da_4}   \right.
\nonumber  \\ & &
\label{4-w}
\hspace{0.1cm}  \left.
                 + \ket{\da_1 ,\da_2, \ua_3, \da_4} +
           \ket{\da_1, \da_2, \da_3, \ua_4} \right)
\end{eqnarray}
in a deterministic way within the framework of cavity QED. The
proposed schemes are based on a bimodal cavity which, in contrast to
single-mode cavities, contains two independent cavity modes (of the
light field). Below, we describe the individual steps of how the
atoms need to interact with either the first or the second cavity
mode, and a graphical language is utilized in order to display these
steps in terms of a quantum circuit. A resonant strong-coupling
regime is assumed, in which the dissipation of the light field in
the cavity is negligible in the course of interaction. After
completion of these steps (the longitudinal experiment), an
entangled GHZ or W state is produced for a chain of four (two-level)
atoms that have passed through the cavity. In practise, however, the
final state of the atoms might not be \textit{pure} but rather a
statistical mixture of states due to decoherence and other
`imperfections' in a given ex\-pe\-ri\-ment. To understand the final
state that is obtained for the atomic chain, we also suggest --- as
the transversal part of the measurements --- a scheme for analyzing
its non-classical correlations, i.e.\ to provide a proof that (or to
which extent) a four-partite GHZ and W state was generated indeed.
The goal is to suggest a scheme that is well adapted to the recent
developments in cavity QED \cite{haroche, haroche1, haroche2} and,
in particular, to the forthcoming generation of high-finesse
microwave cavities that was announced recently \cite{apl90} from the
\textit{Laboratoire Kastler Brossel} (ENS). In addition, we also
show how the scheme below can be generalized quite easily to produce
entangled GHZ and W states for any chain of $N$ two-level Rydberg
atoms.

The paper is organized as follows. In the next Section, we briefly
recall how the resonant interaction of a two-level Rydberg atom with
a given mode of a cavity is described by the Jaynes-Cummings
Hamiltonian, both for single and bimodal cavities. In Sections II.A,
we then present and explain the steps for generating with a chain of
(four) atoms a four-partite GHZ state, and in Sections II.B those
for a W state. For both states, the overall time evolution of the
atoms-cavity system is displayed also in terms of quantum circuits.
These steps are generalized in Section II.C for $N-$partite states,
i.e.\ any number of Rydberg atoms in the chain. In section III,
later, possible set-ups are discussed for performing `transversal'
measurements in order to reveal the  non-classical correlations
within the entangled atom chains. Finally, our conclusions are given
in Section IV.

\begin{figure}
\begin{center}
\includegraphics[width=0.45\textwidth]{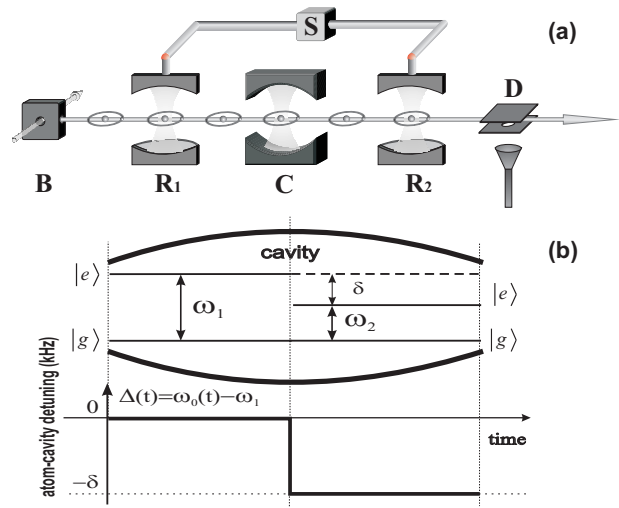} \\
\caption{(a) Schematic set-up of the cavity experiments
\cite{haroche, haroche1} in which a chain of Rydberg atoms from a
source $B$ passes through a Ramsey zone $R_1$, a superconducting
cavity $C$ and the (second) Ramsey zone $R_2$ before the atoms are
field-ionized at the detector $D$. The classical field in the Ramsey
zones is generated by a microwave source $S$. (b) Temporal matching
of the $e \leftrightarrow g$ atomic transition frequency
($\omega_0$) to the frequency $\omega_1$ of the first cavity mode
and the frequency $\omega_2$ of the second mode in the course of the
resonant atomic-cavity interaction. Apart from the matching of the
atomic frequency (upper half), the lower part of this figure
displays the time dependence of the atom-cavity detuning $\Delta(t)
= \omega_{0}(t) - \omega_{1}$, implying a step-wise change from the
resonant $A-C_1$ interaction regime to the resonant $A-C_2$ regime.
See text for further discussions.}
\label{fig:1}
\end{center}
\end{figure}

%
%atom-cavity
%
%
%
\section{Engineering of entangled states by using bimodal cavities}

The \textit{resonant} atom-cavity interaction regime is perhaps the
simplest way to entangle in a controlled manner the atomic circular
states and the quantized cavity field states with each other. For a
sufficiently high quality (factor) of the cavity mirrors, this
regime implies a `strong' atom-field coupling for which the
dissipation of field energy in course of the atom-cavity interaction
becomes negligible. Indeed, avoiding the dissipation of the cavity
field is crucial for engineering multi-partite entangled states of
atomic and/or photonic qubits in a deterministic way. Beside of the
quality of the cavity, the correct matching of the atomic frequency
to the frequency of cavity mode (the so-called \textit{detuning}) is
also important in order to achieve a resonant interaction regime.

In the following, let us adopt the language of Haroche and coworkers
\cite{haroche} (see also \cite{haroche1, haroche2}) for describing
cavity QED experiments and to specify the circular states of the
atoms and the state of cavity. In their experiments, rubidium atoms
are prepared to occupy one of the three (circular) levels with
principal quantum numbers 51, 50, and 49 to which they refer as
exited state $\ket{e}$, ground state $\ket{g}$, and state $\ket{i}$,
respectively. Owing to the design of the cavity \cite{sc288},
however, only the states $\ket{e}$ and $\ket{g}$ can be involved in
the atom-cavity interaction because only the $e \leftrightarrow g$
transition frequency of the Rydberg atoms can be tuned to the
frequency of the cavity mode(s). The classical field from the
microwave source S [see Fig.~\ref{fig:1}(a)], in contrast, can be
adopted to drive either the $e \leftrightarrow g$ or $g
\leftrightarrow i$ transitions and is utilized for generating
superpositions between these states.

The (time) evolution of an atom with a single-mode cavity is
described, both for a resonant and non-resonant interaction, by the
Jaynes-Cummings Hamiltonian \cite{jc} ($\hbar = 1$)
\begin{equation}\label{ham}
H = \omega_{0} S_{z} - \im \frac{\Omega}{2}
        \left( S_{+} a_1 - a_1^{+} S_{-} \right)
      + \omega_1 \left(a_1^{+} a_1 + \frac{1}{2} \right),
\end{equation}
where $\omega_{0}$ is the atomic $e \leftrightarrow g$ transition
frequency, $\omega_1$ the frequency of the cavity field, and
$\Omega$ the atom-field coupling frequency. In this Hamiltonian,
moreover, $a_1$  and $a_1^+$ denote the annihilation and creation
operators for a photon in the cavity, which act upon the Fock states
$\ket{n}$, while $S_{-}$ and $S_{+}$ are the atomic spin lowering
and raising operators that act upon the states $\ket{e}$ and
$\ket{g}$, where the atomic states $\ket{e}$ and $\ket{g}$ are
treated as the `eigenstates' of the spin operator $S_z$ with
eigenvalues $+1/2$ and $-1/2$, respectively. If there is not more
than \textit{one} photon in the cavity, then the overall atom-field
state evolves during the resonant atom-cavity interaction, e.g., for
a zero detuning ($0 = \omega_0 - \omega_1$), as
\begin{subequations}\label{e1}
\begin{eqnarray}
\ket{e,0} & \rightarrow &
\cos{\left(\Omega t /2 \right)} \,\ket{e,0} +
\sin{\left(\Omega t /2 \right)} \,\ket{g,1}, \\[0.1cm]
\ket{g,1} & \rightarrow &
\cos{\left(\Omega t /2 \right)} \,\ket{g,1} -
\sin{\left(\Omega t /2 \right)} \,\ket{e,0} \, ,
\end{eqnarray}
\end{subequations}
i.e.\ with a time evolution that is known also as Rabi rotation. In
this `rotation', $t$ is the effective atom-cavity interaction time
in the laboratory, $\Omega \cdot t$ the respective angle, and a
coupling constant $\Omega/ 2 \pi = 47$ kHz has been utilized in
various cavity QED experiments \cite{sc288, haroche, haroche1,
haroche2}. Note that neither the state $\ket{e,1}$ nor $\ket{g,0}$
appears in the time evolution (\ref{e1}) in line with our physical
intuition that the photon energy is `stored' either by the atom or
the cavity but cannot occur twice in the system.

In order to minimize the contribution of thermal photons, which
occur in microwave cavities due to thermal field leaks, the cavity
is cooled down to $0.6 \, K$ in the experiments \cite{haroche,
haroche1, haroche2}. Moreover, at the beginning of each experimental
sequence the thermal photons are further minimized by sending an
atom in its ground state through the cavity so that it interacts
with a cavity mode for a $\pi$ Rabi rotation and thus `absorbs' the
remaining thermal photons from the cavity mode. By making use of
both, such cooling and `erasing' techniques, an average number of
$n_{\rm\, th} \simeq 0.02$ thermal photons has been achieved so far.
This ensures that the destructive contribution of thermal photons on
the evolution of cavity states during the main experimental sequence
can be neglected.

In contrast to single-mode cavities, a bimodal cavity supports two
independent and non-degenerate modes of light with different
(orthogonal) polarization. Since the frequencies of these light
modes are fixed by the geometry of the cavity, the atomic $e
\leftrightarrow g$ frequency need to be tuned in order that the atom
interacts resonantly with either the first \textit{or} the second
field mode \cite{pra64}. In the language of quantum information, the
additional cavity mode gives rise to another photonic qubit that may
interact independently with the atomic qubits that pass through the
cavity. Indeed, the design and development of bimodal cavities has
been found an important step towards the coherent manipulation of
complex quantum states and for performing fundamental tests in
quantum theory \cite{haroche2, jmo51, pra66_1, cp16, pra67, pra68,
pra76, pla339, jpb}. Below, we shall denote the cavity modes by
$C_{1}$ and $C_{2}$ and suppose that they are associated with the
frequencies $\omega_1$ and $\omega_2$, such that $\omega_1 -
\omega_2 \equiv \delta > 0$. Owing to this \textit{fixed} splitting
in the frequency of the field modes, we refer to the detuning of the
atomic frequency with regard to the cavity modes briefly as
\textit{atom-cavity detuning}. For the cavity utilized in the
experiment by Rauschenbeutel and coworkers \cite{pra64} (see also
\cite{haroche, haroche1, haroche2}), especially a frequency
splitting of $\delta / 2\pi = 128.3$ KHz was realized.

An entanglement of a Rydberg atom with the photon field of the
cavity is achieved by tuning the $e \leftrightarrow g$ transition
frequency as function of time from being `in resonance' with one or
the other cavity mode, while the atom passes through the cavity. For
a proper detuning $\Delta(t)$ of the atomic frequency, a resonant
interaction (regime) is then realized and can be switched between
the two field modes. As seen from the lower part of
Figure~\ref{fig:1}(b), the atom is in resonance with the cavity mode
$C_{1}$ for $\Delta(t) = 0$ and with $C_{2}$ for $\Delta(t) =
-\delta$, where a step-wise change from the $A - C_1$ to the $A -
C_2$ resonant interaction is required. In practise, however, this
step-wise change in the detuning $\Delta(t)$ is experimentally not
feasible. In the experiments by Haroche and coworkers, the detuning
is changed by applying a well adjusted time-varying electric field
across the gap between the cavity mirrors, so that the required
(Stark) shift of the atomic transition frequency $\omega_0(t)$ is
achieved. Instead of a sharp `step-wise' change of the atom-cavity
detuning, therefore a rather smooth `switch' is produced within a
finite time $\tau \simeq 1 \, \mu s$ that corresponds to a
$\frac{\pi}{10}$ angle in units of Rabi rotations. For a typical
atom-cavity interaction time, this finite switch is not negligible
and does affect the evolution of the cavity states \cite{jpb}. In
this paper, however, we shall not consider the effects of this
finite switch, but shall assume a step-wise change in the detuning
as indicated in the lower part of Figure~\ref{fig:1}(b). From the
experimental viewpoint, further improvements of the time-varying
electric field characteristics are needed in order to produce a
sufficiently short (and thus negligible) `switching' time from the
$A - C_1$ to the $A - C_2$ resonant interaction.

We also note that, if the atom is tuned into resonance with one of
the cavity modes, the second mode is \textit{frozen out} from the
atom-cavity interaction owing to the (large) splitting $\delta$
between the two cavity modes. Therefore, the overall $A - C_{1} -
C_{2}$ time evolution of the atom-cavity state can be safely
separated into two independent parts: the evolution due to the $A -
C_{1}$ resonant interaction and that due to $A - C_{2}$. In
practise, however, the splitting between the two cavity modes
frequencies is often not large enough (for example, $\delta \approx
3 \, \Omega$ in the experiment of Ref.\ \cite{pra64}), and then
neither one of the two cavity modes can be frozen out completely.
This leads to a simultaneous interaction of the atom with both
cavity modes and yields an effective \textit{mode wave mixing} in
the cavity \cite{jpb}. Again, we shall not consider the simultaneous
interaction with both cavity modes in this paper but assume the
shift $\delta$ to be sufficiently large, so that the atom-cavity
state evolves according to Eq.\ (\ref{e1}) during the $A - C_{1}$
interaction, and according to
\begin{subequations}\label{e2}
\begin{small}
\begin{eqnarray}
\ket{e,\bar{0}} & \rightarrow & e^{i \delta t}
   \left[ \cos{\left(\Omega t /2 \right)} \ket{e,\bar{0}} +
          \im \sin{\left(\Omega t /2 \right)} \ket{g,\bar{1}}
          \right], \quad
\\[0.1cm]
\ket{g,\bar{1}} & \rightarrow & e^{i \delta t}
   \left[ \cos{\left(\Omega t /2 \right)} \ket{g,\bar{1}} +
          \im \sin{\left(\Omega t /2 \right)} \ket{e,\bar{0}}
          \right], \quad
\end{eqnarray}
\end{small}
\end{subequations}
during the $A - C_{2}$ interaction (period). In the evolution
(\ref{e2}), the states $\ket{\bar{0}}$ and $\ket{\bar{1}}$ hereby
refer to the Fock states of the cavity mode $C_2$, the $\im$ factor
arises due to orthogonal polarization of the mode $M_2$ with respect
to mode $M_1$, and the phase factor $e^{i \delta t}$ arises from the
energy difference $\hbar \delta$ between the two cavity modes being
accumulated in the course of a Rabi rotation.

With this short reminder on the Jaynes-Cummings Hamiltonian and the
(atom-cavity) interactions in a bimodal cavity, we are now prepared to present
the steps that are necessary in order to generate four-partite entangled
states for a chain of Rydberg atoms.

\begin{figure}
\begin{center}
\includegraphics[width=0.475\textwidth]{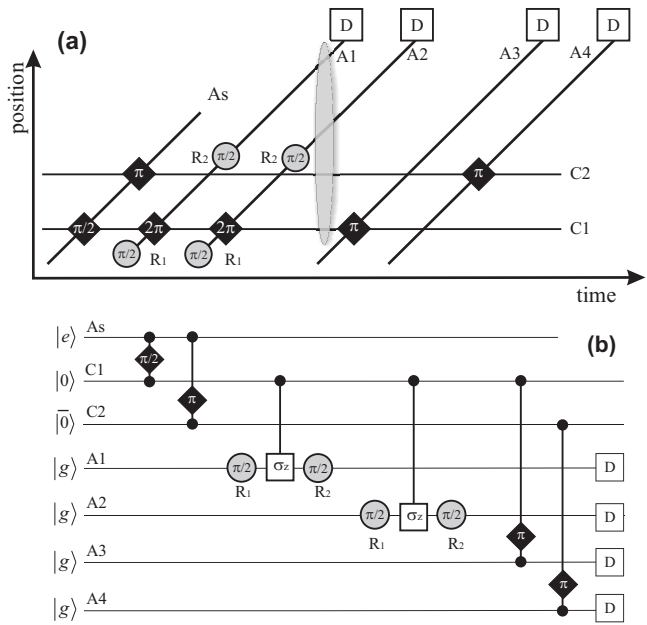}\\
\caption{(a) Temporal sequence for generating a four-partite GHZ
state associated with the chain of Rydberg atoms $A_1$, $A_2$,
$A_3$, and $A_4$. The pictograms in this figures are described in
the text. The gray shadowed ellipse denote the time when the
entanglement of four qubits, the two photonic qubits $C_1$, $C_2$
and the two atomic qubits $A_1$, $A_2$ is achieved. (b) The
corresponding quantum circuit in which the (four) Rydberg atoms are
represented by the four lower lines, being initially prepared in the
ground state $\ket{g}$, while the source atom $A_s$ is shown as the
uppermost line.}
\label{fig:2}
\end{center}
\end{figure}

\subsection{Four-partite GHZ state}

Let us first consider the four-partite GHZ state (\ref{4-ghz}) and
assume that, initially, the cavity is `empty', i.e.\ being in the
state $\ket{0, \bar{0}} \equiv \ket{0} \times \ket{\bar{0}}$. Then,
by using an auxiliary (Rydberg) source atom $A_s$ in the excited
state $\ket{e}$, we can prepare the cavity in a superposition of the
two cavity modes
\begin{equation}
\label{ghz_p_1}
\ket{\Psi_1} =
\frac{1}{\sqrt{2}} \left( \im e^{i \delta \pi / \Omega} \ket{0, \bar{1}}
                        + \ket{1, \bar{0}} \right) \, .
\end{equation}
According to Ref.~\cite{pra64} and our discussions above, this is achieved
if the source atom first interacts with the mode $C_1$ ($\Delta = 0$) for
a Rabi rotation $\Omega \, t_0 = \pi/2$, and afterwards with the cavity mode
$C_2$ ($\Delta = -\delta$) for the rotation $\Omega \, t_1 = \pi$. Owing to
the `rotations' of the atom-cavity state in Eqs.\ (\ref{e1}) and (\ref{e2}),
we shall briefly refer to these interactions as Rabi $\pi/2$, respectively,
$\pi$ pulse, and display them in the Figures by means of black diamonds with
the rotation angle indicated inside. For a resonant interaction, of course,
the subsequent application of these two rotations in Eqs.~(\ref{e1})-(\ref{e2})
with regard to the field modes $C_1$ and $C_2$ leads to a factorization of
the source atom in its ground state $\ket{g}$, and that is therefore omitted
from our further discussion (for further details, see Ref.~\cite{pra64} where
this two-step sequence has been demonstrated also experimentally). For the
sake of brevity, moreover, we shall not display explicitly the values $\Delta$
for the detuning of the atomic frequency which can easily be read off from
the cavity modes as involved in some particular step of the resonant
interaction.

In addition to the resonant atom-cavity interaction, we need to
consider also the interaction of the Rydberg atoms with a
(classical) microwave field that gives rise to a (coherent)
superposition of atomic states before or after they pass through the
cavity, and in dependence of the microwave pulse duration and its
frequency (which can be tuned to the $e \leftrightarrow g$ or $g
\leftrightarrow i$ atomic transitions). In the literature, such an
interaction with a classical field is often called a Ramsey pulse
and is denoted in the Figures by grey circles, showing the
interaction time in units of Ramsey rotations. In addition, we shall
associate the letters $R_1$ or $R_2$ to these circles in order to
denote the Ramsey zone in front or behind the cavity, see
Fig.~\ref{fig:1}(a).

To generate a GHZ state for a chain of Rydberg atoms, being
initially in the ground states $\ket{g}$, we can proceed as follows.
If the cavity is in the superposition (\ref{ghz_p_1}), the state of
atom $A_1$ is first transformed just before it enters the cavity to
\begin{equation} \label{transf-g_1}
\ket{g_1} \rightarrow \frac{1}{\sqrt{2}} (\ket{i_1} + \ket{g_1}) \, .
\end{equation}
This is achieved by using a $\pi/2$ Ramsey pulse tuned to the
$g \leftrightarrow i$ transition frequency while the atom $A_1$ crosses the
first zone $R_1$. For the sake of brevity, we shall denote these interactions
by $R_1(\pi/2, \omega_{g \leftrightarrow i})$, with $\omega_*$ being the
frequency of the microwave source $S$. After the atom $A_1$ has left the Ramsey
zone, it enters the cavity and interacts with mode $C_1$ for a Rabi rotation
$\Omega \, t_3 = 2\pi$. The overall atom-cavity state then becomes

\begin{small}
\begin{equation} \label{ghz_p_3}
\ket{\Psi_3} =
\frac{1}{2} \left( \im e^{i \delta \frac{3\pi}{\Omega}}
                   \ket{ (g_1 + i_1), 0, \bar{1}}
         - \ket{(g_1 - i_1), 1, \bar{0}} \right) \, .
\end{equation}
\end{small}

\noindent The effect of this $2\pi$ rotation can be seen easily from
Eqs.~(\ref{e1}) which implies that the transformation $\,\ket{e_1,0}
\rightarrow -\ket{e_1,0}$ and $\ket{g_1,1} \rightarrow
-\ket{g_1,1}\,$ is made. Of course, this transformation just
describes a $\sigma_z = 2 S_z$ quantum (logic) gate that is applied
to the atom-cavity system. After the atom has passed through the
cavity, it is subjected again to a $R_2(\pi/2,\omega_{g
\leftrightarrow i})$ pulse inside the second Ramsey zone [see
Fig.~\ref{fig:2}(a)], thus leading to the state
\begin{equation}\label{ghz_p_4}
\ket{\Psi_4} =
\frac{1}{\sqrt{2}} \left( \im e^{i \delta 3\pi  / \Omega}
                          \ket{ i_1, 0, \bar{1}}
            - \ket{g_1, 1, \bar{0}} \right) \, ,
\end{equation}
and where the unitary transformation [cf.\ (\ref{transf-g_1})]
\begin{equation} \label{transf}
\ket{g} \rightarrow \frac{1}{\sqrt{2}} (\ket{i} + \ket{g}) \, ,
\quad \ket{i} \rightarrow \frac{1}{\sqrt{2}} (\ket{i} - \ket{g}) \,
\end{equation}
has been utilized. A more detailed discussion of these (three) steps
was given in Ref.~\cite{sc288}, where this sequence of Ramsey and
Rabi pulses was demonstrated also experimentally for the first time.
After the atom $A_1$, the second atom $A_2$ from the chain undergoes
the same temporal sequence of interactions. Leaving apart the
details, these transformations results in the state
\begin{equation} \label{ghz_p_7}
\ket{\Psi_7} =
\frac{1}{\sqrt{2}} \left( \im e^{i \delta 5\pi  / \Omega}
                          \ket{ i_1, i_2, 0, \bar{1}}
            + \ket{g_1, g_2, 1, \bar{0}}
\right) \, ,
\end{equation}
after the second atom has left the set-up. Let us note that already
now we have generated a four-partite GHZ type state for the two
atoms $A_1$ and $A_2$ as well as the two cavity modes $C_1$ and
$C_2$, respectively. For the atoms, moreover, only the two neighbor
states $\ket{i}$ and $\ket{g}$ are involved in the expression
(\ref{ghz_p_7}). In order to generate the GHZ state for a chain of
four atoms, we need to map the information of the photonic qubits
upon the Rydberg atoms $A_3$ and $A_4$. This is done quite easily if
the atom $A_3$ interact with the mode $C_1$ for $\Omega \, t_8 =
\pi$ and atom $A_4$ with the mode $C_2$ for $\Omega \, t_9 = \pi$.
Using Eqs.\ (\ref{e1})-(\ref{e2}), we then see that the cavity
states $\ket{0}$ and $\ket{\bar{0}}$ are mapped upon the ground
states $\ket{g_3}$ and $\ket{g_4}$ of the two atoms, while $\ket{1}$
and $\ket{\bar{1}}$ are mapped upon the exited states $\ket{e_3}$
and $\ket{e_4}$, respectively. For these reasons, the overall
atom-cavity state (\ref{ghz_p_7}) is mapped upon the four Rydberg
atom state
\begin{equation}\label{ghz_p_end}
\ket{\Psi_{\rm\, GHZ}^{(4)}} =
\frac{1}{\sqrt{2}} \left( e^{i \psi} \ket{i_1, i_2, g_3, e_4}
                        + \ket{g_1, g_2, e_3, g_4} \right) \, ,
\end{equation}
whereas the cavity state is factorized out in the vacuum state
$\ket{0, \bar{0}}$. Obviously, the state (\ref{ghz_p_end}) is
equivalent to the state (\ref{4-ghz}) under the change of notation
\cite{notice}
\begin{subequations}\label{ghz_basis}
\begin{eqnarray}
&& \hspace*{-0.8cm}
   \keta{\da_1} = \ket{i_1}, \; \keta{\ua_1} = \ket{g_1}, \;
   \keta{\da_2} = \ket{i_2}, \; \keta{\ua_2} = \ket{g_2},  \\
&& \hspace*{-0.8cm}
   \keta{\da_3} = \ket{g_3}, \; \keta{\ua_3} = \ket{e_3}, \;
   \keta{\da_4} = \ket{e_4}, \; \keta{\ua_4} = \ket{g_4},
\end{eqnarray}
\end{subequations}
except for the factor $e^{i \psi}$, with $\psi =
\frac{7\pi}{\Omega}\delta$, that has no effect on the final-state
probability to find the wave-packet of atomic chain $A_1 - A_4$ in
either the state $\ket{i_1, i_2, g_2, e_4}$ or $\ket{g_1, g_2, e_4,
g_4}$. These probabilities are measured by the detectors, which are
indicated in the figures by the capital $D$ (within a box).

Beside of displaying the individual interactions bet\-ween the atoms
and cavity, that is the particular sequence of Ramsey and Rabi
pulses, a quantum circuit representation of the overall (unitary)
transformation is shown in Fig.~\ref{fig:2}(b). Of course, both
representations (a) and (b) in Fig.~\ref{fig:2} are equivalent and
can be utilized on purpose, where the latter one can be easily
`translated' into quantum gates \cite{nc}. Instead of the $2\pi$
Rabi rotation, the equivalent $\sigma_z$ gate and the initial state
of all (atomic and photonic) qubits are then shown explicitly. This
compact notation for describing the unitary evolution of the
atom-field system in the framework of cavity QED has been introduced
originally by Haroche and coworkers \cite{haroche2} and has been
adopted here for the present discussion.

\begin{figure}
\begin{center}
\includegraphics[width=0.475\textwidth]{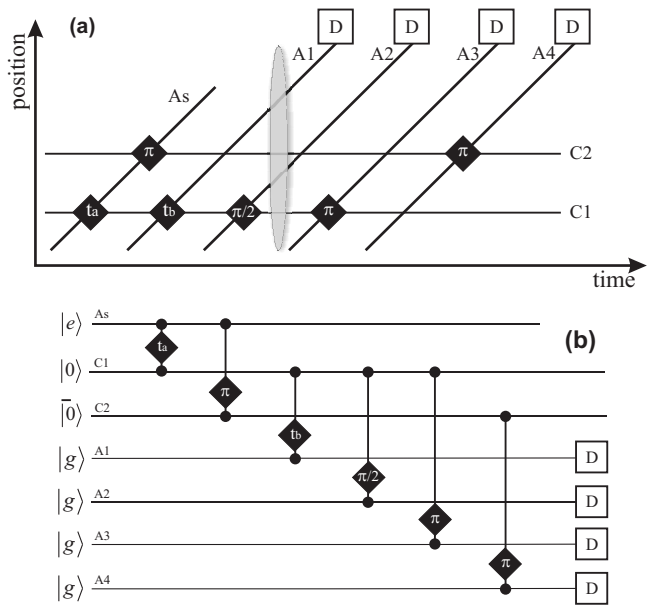}\\
\caption{(a) Temporal sequence for generating a four-partite W state
associated with the chain of Rydberg atoms $A_1$, $A_2$, $A_3$, and
$A_4$. Again, the gray shadowed ellipse denote the time when the
entanglement of four qubits, the two photonic qubits $C_1$, $C_2$
and the two atomic qubits $A_1$, $A_2$ is achieved. (b) The
corresponding quantum circuit in which the (four) Rydberg atoms are
represented by the four lower lines.}
\label{fig:3}
\end{center}
\end{figure}

\subsection{Four-partite W State}

A similar pulse sequence as above for the GHZ state can be worked
out in order to generate a four-partite W state (\ref{4-w}) for a
chain of (four) Rydberg atoms; this pulse sequences can be expressed
again either as temporal sequence for the passage of atoms through
the Ramsey zones and cavity [Fig.~\ref{fig:3}(a)] or as quantum
circuit [Fig.~\ref{fig:3}(b)]. Unlike to the generation of the state
(\ref{ghz_p_1}), however, here we initially prepare the two field
modes of the cavity in the superposition
\begin{equation}\label{w_p_1}
\ket{\Phi_1} =
\frac{1}{2} \left( \im e^{i \delta \pi  / \Omega} \ket{0, \bar{1}}
                 + \sqrt{3} \ket{1, \bar{0}} \right)
\end{equation}
by using the source atom $A_s$ in the exited state $\ket{e}$ and by interacting
first with mode $C_1$ for a Rabi rotation
$\Omega \, t_0 = 2 \arccos{\left( \frac{1}{2} \right)}$ and subsequently with
mode $C_2$ for $\Omega \, t_1 = \pi$. Applying Eqs.~(\ref{e1})-(\ref{e2}),
we see that these pulses result in the cavity state (\ref{w_p_1}), while
the source atom is factorized out in its ground state.

Next, the  Rydberg atoms $A_1$ and $A_2$ pass through the cavity, being
initially in the ground state. We let $A_1$ interact with the mode $C_1$ for
a Rabi rotation $\Omega \, t_2 = 2 \arccos{\left( \sqrt{\frac{2}{3}} \right)}$,
and $A_2$ with $C_1$ for $\Omega \, t_3 = \pi /2$. Then, the overall
atom-cavity state is given by
\begin{eqnarray}\label{w_p_3}
\ket{\Phi_3} & = &
\frac{1}{2} \left( \im e^{i \delta (\frac{3\pi}{2 \Omega} + t_2)}
                   \ket{g_1, g_2, 0, \bar{1}}
         + \ket{g_1, g_2, 1,\bar{0}} \right.
\nonumber \\[0.1cm]
             &   & \hspace{1.4cm}
      \left. - \ket{g_1, e_2, 0, \bar{0}}
             - \ket{e_1, g_2, 0, \bar{0}}\right) \, ,
\end{eqnarray}
i.e.\ by a W-type entangled state between the two photonic qubits
$C_1$, $C_2$ and the atomic qubits $A_1$, $A_2$. To map the
information of the photonic qubits upon the Ryd\-berg atoms $A_3$
and $A_4$, a very similar procedure can be applied as for the GHZ
state in Section II.A: The atom $A_3$ first interacts with mode
$C_1$ for a Rabi rotation $\Omega \, t_4 = \pi$ before $A_4$ enters
the cavity and does the same with $C_2$ for $\Omega \, t_5 = \pi$.
If both atoms enter the cavity in their ground state, the overall
state of the atom chain becomes
\begin{eqnarray}\label{w_p_end}
\ket{\Phi_{\rm W}^{(4)}} & = &
\frac{1}{2} \left( e^{i \phi} \ket{g_1, g_2, g_3, e_4}
                            + \ket{g_1, g_2, e_3, g_4} \right.
\nonumber \\[0.1cm]
             &   & \hspace{0.2cm}
                     \left. + \ket{g_1, e_2, g_3, g_4}
                    + \ket{e_1, g_2, g_3, g_4} \right) \, ,
\end{eqnarray}
while the cavity (state) is factorized out. The state
(\ref{w_p_end}) coincides with the state (\ref{4-w}) under the
change of notation
\begin{subequations}\label{w_basis}
\begin{eqnarray}
&& \hspace*{-0.8cm}
   \keta{\ua_1} = \ket{e_1}, \; \keta{\da_1} = \ket{g_1}, \;
   \keta{\ua_2} = \ket{e_2}, \; \keta{\da_2} = \ket{g_2}, \; \\[0.05cm]
&& \hspace*{-0.8cm}
   \keta{\ua_3} = \ket{e_3}, \; \keta{\da_3} = \ket{g_3}, \;
   \keta{\ua_4} = \ket{e_4}, \; \keta{\da_4} = \ket{g_4}, \;
\end{eqnarray}
\end{subequations}
and where, again, the exponential factor $e^{i \phi}$ with $\phi = $
$ \delta (\frac{7\pi}{2 \Omega} + t_2)$ does not affect the
final-state probability to find the wave-packet of atomic chain $A_1
- A_4$ in either the state $\ket{g_1, g_2, g_3, e_4}$, $\ket{g_1,
g_2 ,e_3, g_4}$, $\ket{g_1 ,e_2, g_3, g_4}$, or $\ket{e_1, g_2, g_3,
g_4}$, respectively.

\subsection{Generation of $N$-partite states}

In the experiments by Rauschenbeutel \etal{} \cite{sc288}, the
generation of the three-partite GHZ state was reported with a
fidelity of $0.54$ \%{}. This rather low value of fidelity, that is
just above of the threshold $1/2$ necessary for proving the
production of this entangled state, is caused mainly by the low
surface quality of the cavity mirrors, i.e.\ the local roughness and
the deviations from the sphe\-ri\-cal geometry, as well as by the
leakage of the cavity field due to its interaction with the
environment. Under the assumption of negligible dissipation, the
`qua\-li\-ty factor' that characterizes the surface quality of
cavity mirrors, is then proportional to the (coherent) photon
storage time and thus determines the number of quantum logical
operations that can be executed successively before the atom-cavity
state becomes completely `destroyed'. This rapid loss of coherence
during the atom-cavity state evolution, has stimulated the group of
\mbox{Raimond} and Haroche at \textit{Laboratoire Kastler Brossel}
(ENS) to develop a new ge\-ne\-ra\-ti\-on of cavity devices that was
announced recently \cite{apl90}. With this new and ultrahigh-finesse
cavity, the `quality factor' was increased by about two orders of
magnitude, in fact, a very remarkable improvement that may enable
them to perform more than hundred quantum lo\-gi\-cal operations
within the `lifetime' of the cavity field. Moreover, by utilizing
the toroidal form of the cavity mirrors (instead of spherical ones
as used previously), the modes frequency splitting $\delta$ was
increased by about one order of magnitude. This large increase
ensures that an atom is coupled to one single mode only and, thus,
that the effective mode wave mixing in the cavity mentioned above
becomes negligible. Owing to this recent success, it seems justified
to suggest new experiments in which multi-partite entangled states
can be generated with a trustworthy fidelity.

\begin{figure}
\begin{center}
\includegraphics[width=0.375\textwidth]{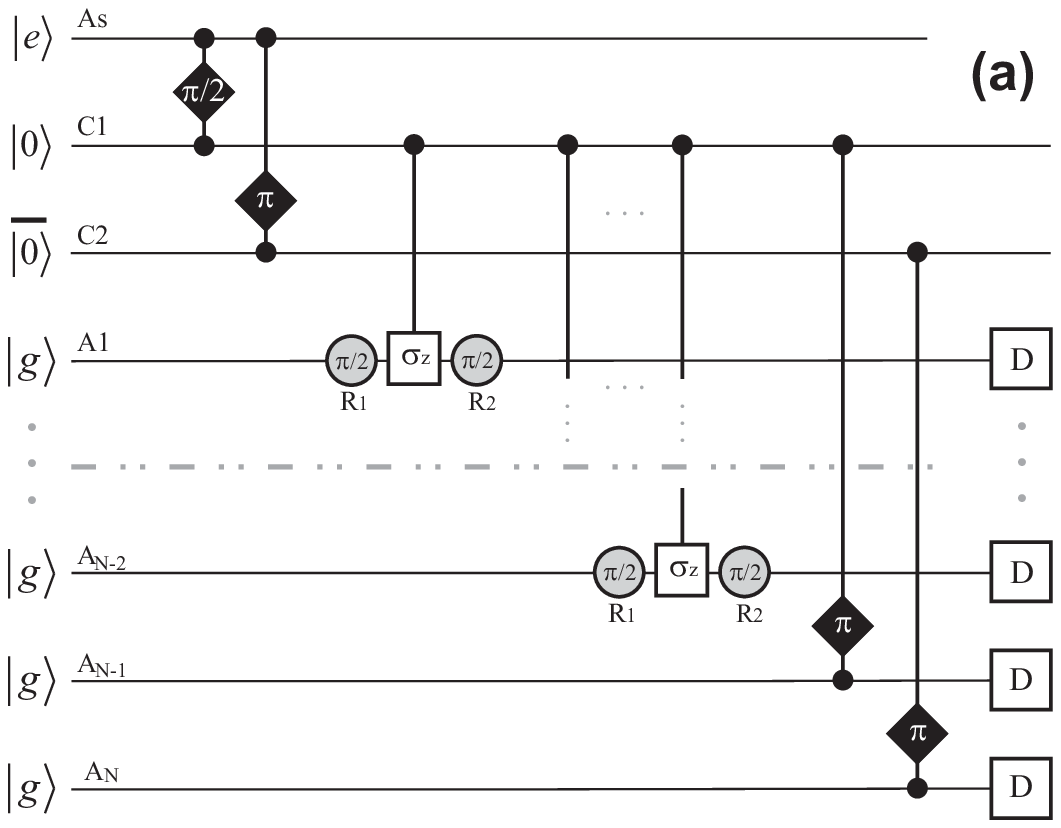}
\vspace{0.8cm}

\includegraphics[width=0.375\textwidth]{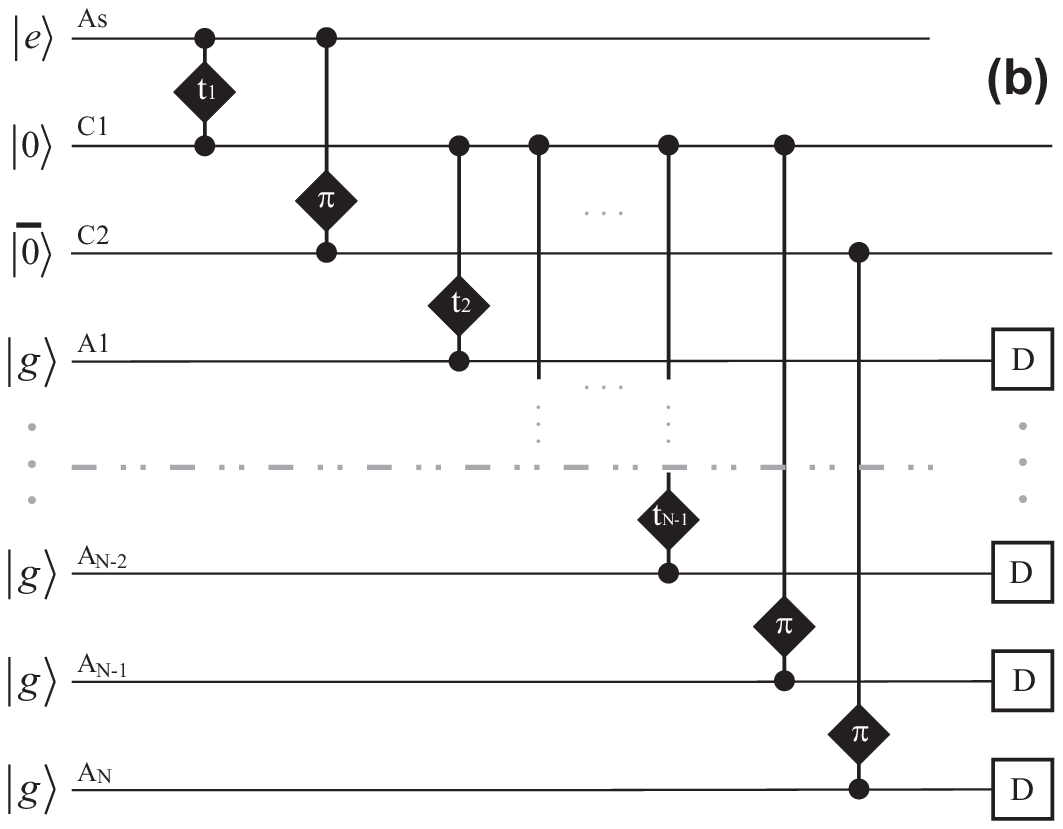}
\caption{(a) Quantum circuit for generating a $N$-partite GHZ state
associated with the Rydberg atoms $A_1 - A_N$. (b) The same but for
the $N$-partite W state; see text for further discussions.}
\label{fig:4}
\end{center}
\end{figure}

To this end, let us consider the $N$-partite (extension to the) GHZ and
W states
\begin{small}
\begin{eqnarray}
\ket{\Psi_{\rm\, GHZ}^{(N)}} & = &
\frac{1}{\sqrt{2}} ( \ket{\ua_1, \ldots, \ua_N} + \ket{\da_1, \ldots, \da_N} ),
\nonumber \\[0.2cm]
\ket{\Phi_{\rm\, W}^{(N)}} & = &
\frac{1}{\sqrt{N}} ( \overbrace{ \ket{\ua_1, \da_2, \ldots, \da_N}
                               + \ldots
                   + \ket{\da_1, \da_2, \ldots, \ua_N}}^{N
                     \rm \;\: terms} ).
\nonumber
\end{eqnarray}
\end{small}

\noindent For these states, $N$-partite entanglement can be
ge\-ne\-ra\-ted in a similar way as discussed above by applying the
individual steps in Fig.~\ref{fig:4}(a) and \ref{fig:4}(b). In this
procedure, a fully entangled state is first generated for $N-2$
atoms and the cavity; then, the information from the cavity field
modes are `mapped' upon $2$ additional Rydberg atoms in order to
obtain a GHZ or W state associated with the atom chain $A_1 - A_N$.
The time intervals $t_i$ to perform the individual Rabi rotations on
the atom-cavity states [Fig.~\ref{fig:4}(b)] are given by
\begin{equation}\label{n-}
t_1 = \frac{2}{\Omega} \arccos{\left(\frac{1}{ \sqrt{N}} \right)};
\; t_n = \frac{2}{\Omega} \arccos{\left( \sqrt{\frac{N-n}{N-n+1}}
\right)} \,  \nonumber
\end{equation}
with $n = 2, \,\ldots, \, N-1$. Not much need to be said here about these
operations since the individual steps can be easily recognized from
Fig.~\ref{fig:4} as well as from our discussion in Sections II.A and B above.

Suppose one could implement the extensions above, the question
naturally arises is up to which $N$ one may proceed in line of the
recent developments in cavity QED. To estimate such a practical
limit in the number of atoms $N$, let us consider the most time
consuming scenario --- the $N$-partite GHZ state, for which each
additional atomic qubit is `incorporated' into the final entangled
state for the price of a $2 \pi$ Rabi rotation ($\sigma_z$ gate). If
we assume that the (minimum) distance between any two successive
atoms is equal to the triple waist length of the cavity mode, then,
the approximate relation between $N$ and the `lifetime` of the
atom-cavity system $T$, takes the form
\begin{equation}\label{relation}
N \simeq \frac{1}{6} \frac{T}{T_{\pi}} \, \varepsilon,
\end{equation}
where $T_{\pi}$ is the required time for a single $\pi$ Rabi
rotation, and $\varepsilon$ a factor which reflects various
corrections to our idealized estimate. Such necessary corrections
might concern the imperfections in the Rabi and Ramsey pulses,
events with two atoms in the same cavity mode, contributions due to
noisy channels, etc. Of course, such additional disturbances can
lead only to a further decrease of $N$, the number of atoms in the
chain. For the atomic velocity $v = 500 \, m/s$, as utilized in the
experiments \cite{haroche, haroche1, haroche2}, a single $\pi$ Rabi
rotation takes about $T_{\pi} \approx 10 \, \mu s$. According to
Ref.\ \cite{apl90}, moreover, the lifetime of the system is bounded
only by the radiative lifetime of the atoms $T \simeq 30 \, ms$ (in
contrast to the cavity photon storage time $\simeq 120 \, ms$).
Using a conservative estimate of $\varepsilon = 0.2$ in
Eq.~(\ref{relation}), then, the number of atoms which may pass the
cavity within the above lifetime $T$ is given by $N \simeq 100$. In
practise, this number must certainly be re-scaled in accordance with
the physical distances between the atomic source, cavity, and
detectors as utilized in a particular experiment.

\section{Detection of the Four-partite GHZ and W States}

Each scheme for generating experimentally a particular (entangled)
multi-partite state for a given atom chain should come along with a
`recipe' that enables one to `demonstrate' that the requested state
has indeed been produced. Since, up to the present, we were mainly
concerned with the (Rabi and Ramsey) rotations that are necessary in
order to achieve the desired state entanglement, not much was said
about the detectors $D$ displayed in Figs.~\ref{fig:2}-\ref{fig:3}.
To project the state of a Rydberg atom upon one of its (allowed)
levels $e$, $g$, or $i$, a field ionization technique (detector) is
applied in the experiments \cite{haroche, haroche1}. From the
detector signal, as taken for many chains of Rydberg atoms, then the
probabilities $P_n(e)$, $P_n(g)$, and $P_n(i)$ are deduced for
occupying a particular level. In the experiments, the field
ionization of some atom from the chain is often characterized in the
literature as `longitudinal' measurement (experiment).

To better understand why one distinct projective measurement -- the
`transversal' measurement need to be carried out, let us re-consider
the GHZ state (\ref{ghz_p_end}) from Sections II.A. With probability
$1/2$, we expect to find the atomic chain $A_1 - A_4$ either in the
(basis) state $\ket{i_1, i_2, g_3, e_4}$ or $\ket{g_1, g_2, e_3,
g_4}$, and similarly the probability $1/4$ to find the W state
(\ref{w_p_end}) in one the four (basis) states $\ket{g_1, g_2, g_3,
e_4}$, $\ket{g_1, g_2, e_3, g_4}$, $\ket{g_1, e_2, g_3, g_4}$, and
$\ket{e_1, g_2, g_3, g_4}$, respectively. However, the same
probabilities are obtained also for the (uncorrelated) statistical
mixture of the corresponding basis states, for instance the mixed
state $[ \{1/2,\, \ket{i_1, i_2, g_3, e_4} \}, \:
   \{1/2,\, \ket{g_1, g_2, e_3, g_4} \} ] $.
Therefore, no (longitudinal) measurement alone is sufficient for
proving the non-classical nature of the correlated atomic chain for
GHZ or W type entanglement, but has to be augmented by additional
measurements. The same can be seen already from the (Bell) state
$\frac{1}{\sqrt{2}}(\keta{\ua_1, \da_2} + \keta{\da_1, \ua_2})$ that
describes a rotation-invariant spin singlet state of two qubits. As
well known for such a singlet state, we shall find the two spins
always in opposite direction for \textit{any} choice of the
quantization axis of the (projective) measurement. In the
literature, this counter-intuitive result is known also as
Einstein-Rosen-Podolsky (EPR) paradoxon \cite{pr47}, and this
freedom in the choice of the quantization axis can therefore be
exploited to display the non-classical correlations of the generated
GHZ and W states.

Following the work by Hagley \etal{} \cite{prl79}, let us now adopt
the geometrical language of the Bloch sphere in order to introduce a
more quantitative description for the projective measurement in the
framework of cavity QED. Using the Bloch sphere, any single-qubit
state can be represented as a point either \textit{on} the sphere
(pure states) or \textit{within} the sphere (mixed states).
Moreover, the two basis states $\keta{\ua_n}$ and $\keta{\da_n}$ are
taken along $z$ as the quantization axis that crosses the sphere at
the north and south pole, respectively. In this standard
representation of the Bloch sphere, the $x$ axis is defined by the
vectors $\ket{+^x_n} = \frac{1}{\sqrt{2}} (\keta{\ua_n} +
\keta{\da_n})$ and $\ket{-^x_n} = \frac{1}{\sqrt{2}} (\keta{\ua_n} -
\keta{\da_n})$, the $y$ axis is defined by vectors $\ket{+^y_n} =
\frac{1}{\sqrt{2}} (\keta{\ua_n} + \im \keta{\da_n})$
and $\ket{-^y_n} = \frac{1}{\sqrt{2}} (\keta{\ua_n} - \im
\keta{\da_n})$. In addition, any other axis $\xi(\varphi)$ in the
equatorial $x-y$ plane, that forms the angle $\varphi$ with respect
to the $x$ axis, can be characterized by the unit vectors
$\ket{+^\varphi_n} = \frac{1}{\sqrt{2}} (\keta{\ua_n} + e^{i
\varphi} \keta{\da_n})$ and $\ket{-^\varphi_n} = \frac{1}{\sqrt{2}}
(\keta{\ua_n} - e^{i \varphi} \keta{\da_n})$ where, again, the `+'
and `--' sign is chosen to distinguish between positive and negative
values along the axis. Recall that the basis states $\keta{\ua_n}$
and $\keta{\da_n}$ are related to the two neighbor atomic states
$\ket{e_n}$ and $\ket{g_n}$, or $\ket{g_n}$ and $\ket{i_n}$ via
expressions (\ref{ghz_basis}) or (\ref{w_basis}) for the
four-partite GHZ or W entangled chain of atoms, respectively. By
this choice, we thus defined the $z$ axis of the Bloch sphere as
pointing along our `longitudinal' quantization axis that coincides
with the projection measurement performed by the detector.

With this notation, following Hagley \etal{} \cite{prl79} we can now
explain how one and the same detector (as used for projection along
the $z$ axis) can be applied to perform a projection along either
the $x$ or $\xi(\varphi)$ (transversal) axes, respectively. If we
consider an atom in the superpositions $\frac{1}{\sqrt{2}}
(\keta{\ua} + \keta{\da})$ and $\frac{1}{\sqrt{2}} (\keta{\ua} -
\keta{\da})$, then a resonant $\pi/2$ Ramsey pulse between the two
neighbor levels $\keta{\ua} \leftrightarrow \keta{\da}$ implies the
transformations
\begin{equation} \label{transf1}
\frac{1}{\sqrt{2}} (\keta{\ua} + \keta{\da}) \rightarrow \keta{\da},
\quad \frac{1}{\sqrt{2}} (\keta{\ua} - \keta{\da}) \rightarrow
\keta{\ua} \, .
\end{equation}
Leaving the Ramsey zone, the atom enters the detector, where it is
projected either upon the state $\keta{\da}$ or $\keta{\ua}$
corresponding to poles of the Bloch sphere above. The time reversal
of the (unitary) transformation (\ref{transf1}) thus suggests that a
combination of the resonant $\pi/2$ Ramsey pulse followed by a
standard (longitudinal) measurement, can be viewed as a projective
measurement upon the $x$ axis as given by vectors $\ket{\pm^x} =
\frac{1}{\sqrt{2}} (\keta{\ua} \pm \keta{\da})$. Alternately to the
resonant Ramsey pulse, if we perform a pulse $R_2(\pi/2,
\widetilde{\omega})$ with a frequency $\widetilde{\omega}$ that is
slightly shifted with regard to the atomic transition ($\keta{\ua}
\leftrightarrow \keta{\da}$) frequency $\omega$, then a phase
difference $\varphi = \tau \cdot ( \widetilde{\omega} - \omega)$ is
accumulated by the atomic state during the coherence time $\tau$.
Therefore, a combination of a near-resonant Ramsey pulse with a
tunable frequency $\widetilde{\omega}$ ($\sim \varphi$) followed by
a detection of the atom within the longitudinal basis, is equivalent
to a projective measurement upon the $\xi(\varphi)$ axis as
described by vectors $\ket{\pm^\varphi} = \frac{1}{\sqrt{2}}
(\keta{\ua_n} \pm e^{i \varphi} \keta{\da_n})$. Further details
concerning the transversal measurement in cavity QED can be found in
Refs.~\cite{prl79, sc288}.

\begin{figure}
\begin{center}
\includegraphics[width=0.385\textwidth]{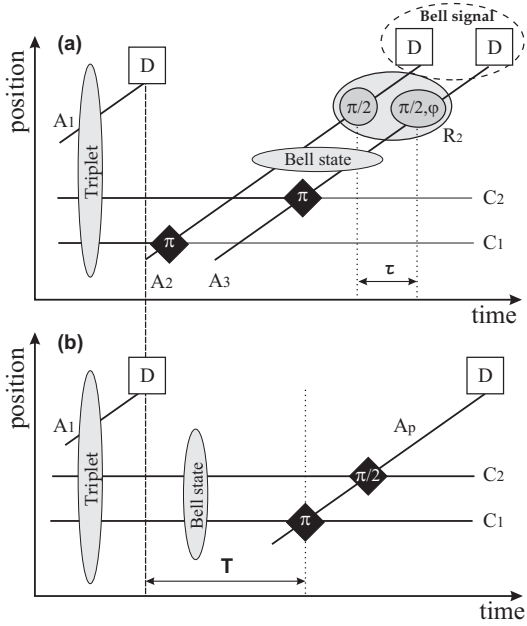}\\
\caption{Sequence (a) displays the basic steps of the transversal
measurement for the GHZ state (`triplet' $A_1 - C_1 - C_2$) given by
expression (\ref{ghz_p_3}). See the text for explanations. Sequence
(b) displays the principle of the another type of measurement as
applied over the same entangled triplet (\ref{ghz_p_3}).}
\label{fig:5}
\end{center}
\end{figure}

To make the above statements clear, let us consider the
three-partite GHZ state (\ref{ghz_p_3}) from Section II.A and
suppose that atom $A_1$ is projected onto the states $\ket{i_1}$ or
$\ket{g_1}$ after it has passed the cavity. For the field modes
$C_1$ and $C_2$, this projection gives rise to a collapse of the
wave-packet (\ref{ghz_p_3}) into one of the (two) Bell states
\begin{equation}\label{ghz_d_1}
\ket{\Psi_{coll}^{\pm}} = \frac{1}{\sqrt{2}} \left( \im e^{i \delta
\frac{3 \pi}{\Omega}} \ket{0, \bar{1}} \pm \ket{1, \bar{0}} \right),
\end{equation}
where the $+$ sign is associated with the atom in the state $\ket{i_1}$ and
the $-$ sign with $\ket{g_1}$, respectively. These Bell states can be mapped
upon the atoms $A_2$ and $A_3$ by following the procedure from Sections II
and as seen in Fig.~\ref{fig:5}(a). After this map, the cavity states are
factorized out and the state of the atoms $A_2$ and $A_3$ then becomes
\begin{equation}\label{ghz_d_2}
   \ket{\widetilde{\Psi}_{coll}^{\pm}}
   = \frac{1}{\sqrt{2}}
   \left( e^{i \eta} \ket{g_2, e_3} \pm \ket{e_2, g_3} \right)
\end{equation}
with $ \eta = 5 \pi \cdot \delta / \Omega$. To perform an
independent (transversal) measurement on these atoms, we project
$A_2$ upon the $x$ and $A_3$ upon the $\xi(\varphi)$ axis. This is
done by acting with a $R_2(\pi/2, \omega_{e \leftrightarrow g})$
pulse on atom $A_2$, followed by a projective measurement in the
longitudinal basis, together with a near-resonant $R_2(\pi/2,
\widetilde{\omega})$ pulse upon $A_3$ as well followed by a
projective measurement in the longitudinal basis [see
Fig.~\ref{fig:5}(a)]. Since the latter Ramsey pulse is done with the
(near-resonant) frequency $\widetilde{\omega}$, the phase difference
$\varphi$ ($\sim \widetilde{\omega}$) is accumulated during the time
$\tau$ given by delay between the $R_2(\pi/2, \omega_{e
\leftrightarrow g})$ and $R_2(\pi/2, \widetilde{\omega})$ pulses.

The above two sequences: (i) $R_2(\pi/2, \omega_{e \leftrightarrow
g})$ acting upon $A_2$ followed by $D$  and (ii) $R_2(\pi/2,
\widetilde{\omega})$ acting upon $A_3$ followed by $D$, together
with the four possible projections $\ket{e_{2,3}}$ and
$\ket{g_{2,3}}$ of the atoms $A_2$ and $A_3$, give eight outcomes of
the measurements with non-zero probability
\begin{subequations}\label{prob_set}
\begin{eqnarray}
\label{prob_set1} P_{\pm}(e_2, e_3; \varphi) = |\left( \bra{-_1^x}
\times \bra{
-_2^\varphi} \right) \ket{\widetilde{\Psi}^{\pm}_{coll}}|^2, && \\[0.1cm]
P_{\pm}(g_2, g_3; \varphi) = |\left( \bra{+_1^x} \times \bra{
+_2^\varphi} \right) \ket{\widetilde{\Psi}^{\pm}_{coll}}|^2, && \\[0.1cm]
P_{\pm}(e_2, g_3; \varphi) = |\left( \bra{-_1^x} \times \bra{
+_2^\varphi} \right) \ket{\widetilde{\Psi}^{\pm}_{coll}}|^2, && \\[0.1cm]
\label{prob_set4}  P_{\pm}(g_2, e_3; \varphi) = |\left( \bra{+_1^x}
\times \bra{ -_2^\varphi} \right)
\ket{\widetilde{\Psi}^{\pm}_{coll}}|^2 . &&
\end{eqnarray}
\end{subequations}
Of course, these probabilities depend (parametrically) on the angle
$\varphi$, that defines the axis for the \textit{transversal}
measurements in the $x-y$ plane, while the subscript $\pm$ refers to
the particular state (\ref{ghz_d_2}) that was obtained after the
projection of atom $A_1$ upon $z$ axis. The probabilities
(\ref{prob_set1})-(\ref{prob_set4}), corresponding to all possible
outcomes for $A_2$ and $A_3$, are then combined for many instances
of one and the same experiment, in order to produce the (so-called)
`Bell signal' \cite{haroche, haroche2, sc288}
\begin{eqnarray}\label{signal}
   I_{\pm}(\varphi) &=&
   P_{\pm}(e_2, e_3; \varphi) + P_{\pm}(g_2, g_3; \varphi)
   \nonumber \\
   && \quad - P_{\pm}(e_2, g_3; \varphi) - P_{\pm}(g_2, e_3; \varphi)
\end{eqnarray}
for any angle $\varphi$ in the interval $[ 0, 2\pi ]$.

For an idealized set-up of the experiment, the signal (\ref{signal})
has the form $I_{\pm}(\varphi) = \pm \cos(\varphi + \eta)$ and thus,
the \mbox{observed} oscillation of the signal as function of
$\varphi$, would reveal non-classical correlations of the Bell
states (\ref{ghz_d_2}). Indeed, the above `recipe' meets the
mentioned request for carrying out an additional measurement upon an
independent quantization axis ($x$ and $\xi(\varphi)$ axes in this
case), and moreover, since the $+$ sign is associated with the atom
$A_1$ being in the state $\ket{i_1}$ and the $-$ sign with
$\ket{g_1}$, this technique of `transversal measurements' enables
one to reconstruct quantum correlations of the initial triplet state
(\ref{ghz_p_3}), see Ref.~\cite{sc288} for details.

Beside of varying the angle $\varphi$, i.e.\ the shift in the
frequency $\widetilde{\omega}$ of the Ramsey pulse $R_2(\pi/2,
\widetilde{\omega})$ with regard to the atomic transition frequency
$\omega$, there is another possibility to perform an independent
measurement on the Bell state (\ref{ghz_d_1}) and which is
particularly suitable for bimodal cavities. This technique is based
on the delay time $T$, that is introduced between the creation of
the Bell state (\ref{ghz_d_1}) among the cavity modes $C_1 - C_2$,
and the time when the cavity state is `probed' by one further atom
$A_p$ [cf.\ Fig.~\ref{fig:5}(b)]. In the latter step, the probe atom
$A_p$ prepared in the ground state, first interacts with the cavity
mode $C_1$ for $\Omega \, t_a = \pi$ and with the mode $C_2$ for
$\Omega \, t_b = \pi/2$, followed by an projection of $A_p$ in the
longitudinal basis. According to Eqs.~(\ref{e1})-(\ref{e2}), for an
idealized experiment, this (two-step) sequence produces the
probability amplitude to detect $A_p$ in the exited state
\begin{equation} \label{prob}
   P_{\pm}(e; \, T) =
   \frac{1 \pm \cos(\delta \cdot T +  4 \pi \delta / \Omega)}{2} \,
   .
\end{equation}
Again, here the $+$ sign is associated with the atom $A_1$ in the
state $\ket{i_1}$ and the $-$ sign with $\ket{g_1}$. An oscillation
of the probability amplitude (\ref{prob}) as function of the delay
time $T$ then proves the coherent superposition of the two cavity
mode states, and thus, provides us with an `entanglement measure'
similarly to the Bell signal (\ref{signal}) in the previous case.
Moreover, the sensibility of this measurement to the sign of the
pair (\ref{ghz_d_1}) enables one also to reconstruct quantum
correlations of the initial triplet state (\ref{ghz_p_3}). Note that
the above `probing' of the cavity modes yields also a non-zero
probability to detect $A_p$ in the ground state that fulfills the
relation $P_{\pm}(g; \, T) = P_{\mp}(e; \, T)$. Therefore, if the
probability (\ref{prob}) is chosen in order to reconstruct the state
(\ref{ghz_p_3}), then one should collect only those probabilities
during the measurement, for which the probe atom has been detected
in its exited state and discard all other events, for which the atom
has been detected in its ground state. Further details concerning
this technique can be found in Ref.~\cite{pra64}.

After these brief explanations of the different types of measurement
techniques for probing non-classical correlations, we are prepared
to discuss those steps which are necessary for analyzing the
four-partite entangled states from Sections~II.A and II.B.

\begin{figure}
\begin{center}
\includegraphics[width=0.475\textwidth]{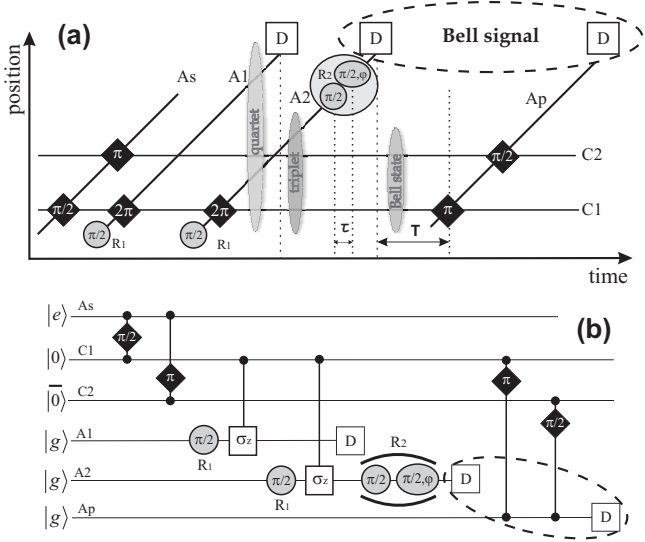}\\
\caption{(a) The temporal sequence for performing transversal
measurements on the four-partite GHZ state (`quartet' $C_1 - C_2 -
A_1 - A_2$) given by expression (\ref{ghz_d_5}). Here the grey
shadowed ellipses: `triplet', and `Bell state' correspond to the
three-partite GHZ state ($C_1 - C_2 - A_2$) and two-partite
entangled state ($C_1 - C_2$). (b) The quantum circuit that
corresponds to the temporal sequence shown above.}
\label{fig:6}
\end{center}
\end{figure}

\subsection{Detection of the four-partite GHZ State}

To analyze the four-partite GHZ state from Section~II.A, we shall
combine both, the transversal and longitudinal measurement
techniques from above. Our goal is to recognize if an (uncorrelated)
statistical mixture of the states $\ket{i_1, i_2, g_3, e_4}$ and
$\ket{g_1, g_2, e_3, g_4}$ occurs during the experiment, since it
leads to the same outcome of the projection (upon the $z$ axis) of
individual atoms from the chain as for entangled GHZ state. Let us
start our analysis by considering the three-partite GHZ state
(\ref{ghz_p_3}) from Section II.A
\begin{equation}\label{ghz_d_3}
   \ket{\Psi_3} = \frac{1}{\sqrt{2}}
      \left( \im e^{i \delta \frac{3\pi}{\Omega}} \ket{ +^x_1, 0, \bar{1}}
                                                - \ket{ -^x_1, 1, \bar{0}}
      \right) \,
\end{equation}
where the Bloch sphere notation $\ket{\pm^x}$ along with relations
(\ref{ghz_basis}) have been used. After the atom $A_1$ leaves the
cavity, atom $A_2$ prepared in the ground state, is subjected in the
first Ramsey zone to the pulse $R_1(\pi/2, \omega_{g \leftrightarrow
i})$ and then, while passing the cavity, it interacts with the mode
$C_1$ for $\Omega \, t_5 = 2 \pi$ as seen in Fig.~\ref{fig:6}(a).
The atom-cavity wave-packet thus results into the four-partite GHZ
state
\begin{equation} \label{ghz_d_5}
   \ket{\Psi_5} = \frac{1}{\sqrt{2}}
   \left( \im e^{i \delta \frac{5\pi}{\Omega}}
      \ket{ +^x_1, +^x_2, 0, \bar{1}}
    + \ket{ -^x_1, -^x_2, 1, \bar{0}} \right).
\end{equation}
In contrast to Section~II.A, we shall not map the information from
the cavity upon some additional atoms but take the (atom-cavity)
state (\ref{ghz_d_5}) itself for performing the transversal
measurements.

As seen from Figure~\ref{fig:6}(a), the atom $A_1$ leaves the cavity
in either the state $\ket{+^x_1}$ or $\ket{-^x_1}$ and is
`projected' in the detector upon the states $\ket{g_1}$ or
$\ket{i_1}$, respectively. This measurement reduces the state
(\ref{ghz_d_5}) to
\begin{equation} \label{ghz_d_6}
   \ket{\Psi^{\pm}_6} = \frac{1}{\sqrt{2}}
   \left( \im e^{i \delta \frac{5\pi}{\Omega}} \ket{  +^x_2, 0, \bar{1}}
                                           \pm \ket{ -^x_2, 1,\bar{0}}
   \right) \,
\end{equation}
where the `+' sign corresponds to the outcome $\ket{g_1}$ and the
`--' sign to $\ket{i_1}$. Next to $A_1$, atom $A_2$ leaves the
cavity and is subjected to the pulse $R_2(\pi/2, \omega_{g
\leftrightarrow i})$ in the second Ramsey zone, the state
(\ref{ghz_d_6}) thus becomes
\begin{equation}\label{ghz_d_7}
   \ket{\Psi^{\pm}_7} = \frac{1}{\sqrt{2}}
   \left( \im e^{i \delta \frac{5\pi}{\Omega}} \ket{  i_2, 0, \bar{1}}
                                           \pm \ket{ g_2, 1,  \bar{0}}
   \right) \, .
\end{equation}
In typical cavity QED experiments \cite{haroche, haroche1,
haroche2}, a single $\pi/2$ Ramsey pulse takes about $1 \mu s - 2
\mu s$ and implies that the atom $A_2$ is still \textit{inside} of
the Ramsey plates when the required `rotation' of the level
population has been completed. After a short time delay $\tau$, it
is therefore possible to address an additional (near-resonant)
$R_2(\pi/2, \widetilde{\omega})$ pulse upon $A_2$ within the same
Ramsey zone. Finally, leaving the Ramsey plates, the atom $A_2$ is
projected on either $\ket{i_2}$ or $\ket{g_2}$ state inside the
detector [cf.\ Fig.~\ref{fig:6}(a)].

As we explained above, the combination of a near-resonant Ramsey
pulse $R_2(\pi/2, \widetilde{\omega})$ together with the measurement
of $A_2$ in its longitudinal basis is equivalent to a projective
measurement upon the $\xi(\varphi)$ axis, where the angle $\varphi =
\tau \cdot (\widetilde{\omega} - \omega_{g \leftrightarrow i})$ is
accumulated in the course of the time $\tau$ given by delay between the
$R_2(\pi/2, \omega_{g \leftrightarrow i})$ and $R_2(\pi/2,
\widetilde{\omega})$ pulses. After the projection of $A_2$ upon
$\ket{i_2}$ or $\ket{g_2}$ , the (two) cavity modes therefore remain
either in the state
\begin{small}
\begin{subequations}\label{ghz_d_8}
\begin{eqnarray}
   \ket{\Psi^{\pm}_8(i_2; \, \varphi)}
   & \equiv & \sqrt{2} \, \braket{+^\varphi_2}{\Psi^{\pm}_7} \nonumber \\
   & = & \frac{1}{\sqrt{2}}
   \left( \im e^{i (- \varphi + \delta \frac{5\pi}{\Omega} )} \ket{0, \bar{1}}
                                                          \pm \ket{1, \bar{0}}
   \right) \label{ghz_d_8a} \\[0.1cm]
   && \hspace*{-3.3cm} {\rm or} \nonumber \\[0.1cm]
   \ket{\Psi^{\pm}_8(g_2; \, \varphi)}
   & \equiv & \sqrt{2} \, \braket{-^\varphi_2}{\Psi^{\pm}_7}  \nonumber \\
   & = & \frac{1}{\sqrt{2}}
   \left( \im e^{i (- \varphi + \delta \frac{5\pi}{\Omega} )} \ket{0, \bar{1}}
                                                          \mp \ket{ 1,\bar{0}}
   \right) \,  , \label{ghz_d_8b}
\end{eqnarray}
\end{subequations}
\end{small}
respectively. Although Eqs.~(\ref{ghz_d_8}) describe the coherent
superposition of the cavity states, they also contain information
about the initial four-partite state (\ref{ghz_d_5}) due to the
phase (angle) $\varphi$ and as well as due to the sign $\pm$.

\begin{figure}
\begin{center}
\includegraphics[width=0.475\textwidth]{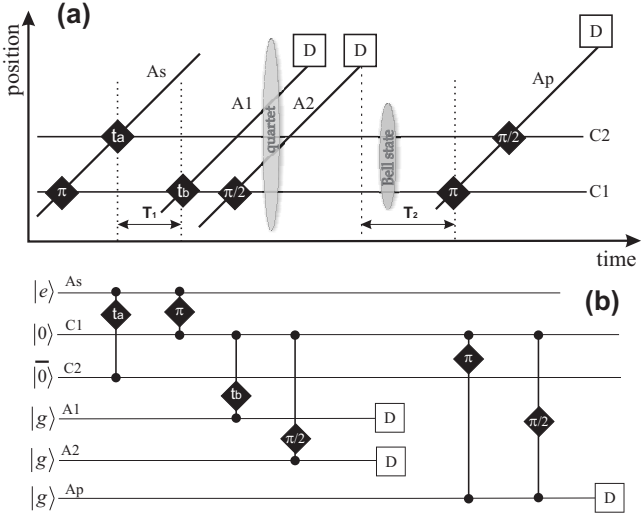}\\
\caption{(a) The temporal sequence for performing transversal
measurements on the four-partite W state (`quartet' $C_1 - C_2 - A_1
- A_2$) given by expression (\ref{w_d_3}). Here the grey shadowed
ellipse `Bell state' corresponds to the two-partite entangled state
($C_1 - C_2$). The time intervals $t_a$ and $t_b$ are given in the
text. (b) The quantum circuit that corresponds to the temporal
sequence shown above.}
\label{fig:7}
\end{center}
\end{figure}

To reveal the entanglement of (\ref{ghz_d_5}) due to measurements on
the state (\ref{ghz_d_8}), we can utilize the last measurement
technique from above [see Fig.~\ref{fig:5}(b)] by introducing a
proper time delay $T$ that contributes to the cavity state by means
of the energy difference $\hbar \delta \, T$ bet\-ween the modes.
During this time delay, the `free' evolution of the cavity state
results in the phase shift $e^{i (- \varphi + 5\pi \delta / \Omega)}
\; \rightarrow \;
  e^{i (T \delta - \varphi + 5\pi \delta / \Omega)}$.
After this time delay, the probe atom $A_p$ enters the cavity and
interacts with the mode $C_1$ for $\Omega \, t_a = \pi$ and with
$C_2$ for $\Omega \, t_b = \pi/2$ as shown in Fig.~\ref{fig:6}(a).
According to Eqs.~(\ref{e1})-(\ref{e2}), the last (two-step)
sequence makes the overall state of the cavity and probe atom to
becomes either
\begin{eqnarray}
   \ket{\Psi^{\pm}_{10}(i_2; \, \varphi,T)}
   = \frac{1}{2} \left[ \im \left( 1 \mp e^{i (T\delta - \varphi + \theta)}
                            \right) \ket{g_p, 0,\bar{1}}
         \right. && \nonumber \\
           + \left.     \left( 1 \pm e^{i (T\delta - \varphi + \theta)}
                        \right)  \ket{e_p, 1, \bar{0}}
             \right] , &&
\end{eqnarray}
if the cavity was initially in the state (\ref{ghz_d_8}a), or
\begin{eqnarray}
   \ket{\Psi^{\pm}_{10}( g_2; \, \varphi,T)}
   = \frac{1}{2} \left[ \im \left( 1 \pm e^{i (T\delta - \varphi + \theta)}
                            \right) \ket{g_p, 0,\bar{1}}
         \right. && \nonumber \\
           + \left.     \left( 1 \mp e^{i (T\delta - \varphi + \theta)}
                        \right) \ket{e_p, 1, \bar{0}}
         \right] \, , &&
\end{eqnarray}
for the state (\ref{ghz_d_8}b), where $\theta = 6 \pi \cdot \delta /
\Omega$. Therefore, the four possible projections $\ket{i_2 /g_2}$
and $\ket{e_p /g_p}$ of the atoms $A_2$ and $A_p$, give the
following eight outcomes of the pro\-ba\-bi\-li\-ty amplitudes
\begin{subequations}
\begin{eqnarray}
   P_{\pm}(i_2, g_p; \, \varphi,T)
   & = & |\braket{g_p}{\Psi^{\pm}_{10}(i_2; \, \varphi, T)} |^2  \\
   P_{\pm}(g_2, e_p; \, \varphi,T)
   & = & |\braket{e_p}{\Psi^{\pm}_{10}(g_2; \, \varphi, T)} |^2  \\
   P_{\pm}(i_2, e_p; \, \varphi,T)
   & = & |\braket{e_p}{\Psi^{\pm}_{10}(i_2; \, \varphi, T)} |^2  \\
   P_{\pm}(g_2, g_p; \, \varphi,T)
   & = & |\braket{g_p}{\Psi^{\pm}_{10}(g_2; \, \varphi, T)} |^2
\end{eqnarray}
\end{subequations}
that depend on the angle $\varphi$ and the time delay $T$, and where
the sign `$\pm$' refers again to the outcome $\ket{g_1}$ or
$\ket{i_1}$ for the first atom $A_1$. As before, these probabilities
are then combined for many instances of the same temporal sequence,
thus producing the correlation signal
\begin{eqnarray}\label{signal1}
   I_{\pm}(\varphi, T) &=&
   P_{\pm}(i_2, g_p; \, \varphi, T) + P_{\pm}(g_2, e_p; \, \varphi, T)
   \nonumber \\ && \hspace*{-0.3cm}
   - P_{\pm}(i_2, e_p; \, \varphi, T) - P_{\pm}(g_2, g_p; \, \varphi,
   T), \quad
\end{eqnarray}
which is obtained during the experiment. For an idealized set-up,
this signal takes the form $I_{\pm}(\varphi, T) = \mp \cos
\left(T\delta - \varphi + \theta \right)$, and where we have the
angle $\varphi$ ($\sim \widetilde{\omega}$) and the time delay $T$
as two independent parameters that can be varied in order to proof
or discard that the desired four-partite state (\ref{ghz_d_5}) was
indeed generated.

\subsection{Detection of the four-partite W State}

The ($N-$partite) GHZ and W state are essentially different in that
they cannot be transformed into each other under any LOCC
operations. For this reason, a quite different (temporal) sequence
of transversal measurements has to be found in order to prove that
the four--partite W state was indeed generated by a given
experimental sequence. To develop such a sequence, let us note that
the four-partite W state (\ref{w_p_3}) can be cast into the form
\begin{eqnarray} \label{w_d_1}
   \ket{\Phi_3} = \frac{1}{\sqrt{2}}
   \left[ \frac{1}{\sqrt{2}}
          \left( \im e^{i \delta (\frac{3\pi}{2 \Omega} + t_2)}
             \ket{0, \bar{1}}
           + \ket{1, \bar{0}}
      \right) \ket{g_1, g_2}
   \right. \hspace{0.3cm} &&
   \notag \\
   \left. - \frac{1}{\sqrt{2}}
          \left( \ket{g_1, e_2} + \ket{e_1, g_2} \right)
      \ket{0, \bar{0}}
   \right]  \, , \hspace{1.0cm} &&
\end{eqnarray}
in which we have two types of (two-partite) Bell states: (i) the
`photonic' Bell state in the first line, and (ii) the `atomic' Bell
state in the second line. This representation of the W state
therefore suggests a temporal sequence in which the photonic Bell
state is `coherently' isolated and for which the entanglement is
shown independent of the atomic part. Indeed, such a (temporal)
sequence is shown Fig.~\ref{fig:7}(a) and is re-drawn as quantum
circuit in Fig.~\ref{fig:7}(b).

Following Fig.~\ref{fig:7}, we start from an empty cavity in the state
$\ket{0, \bar{0}}$, and make use of an auxiliary source atom $A_s$ to
prepare the cavity in the superposition
\begin{equation} \label{w_d_2}
   \ket{\Phi_1} = \frac{1}{2}
   \left( \im \ket{0, \bar{1}}
        + \sqrt{3} \, e^{i \frac{\delta \pi}{\Omega}} \ket{1, \bar{0}}
   \right)  \, .
\end{equation}
This is done, if the atom first interacts with the mode $C_2$ for
$\Omega \, t_a = 2 \arccos{\left( \sqrt{\frac{3}{4}} \right)}$ and,
thereafter, with the mode $C_1$ for $\Omega \, t_1 = \pi$. Before
the atom $A_1$ enters the cavity, we let the field state
(\ref{w_d_2}) evolve freely for the time delay $T_1$, which leads to
the phase shift $e^{i \delta \pi / \Omega} \, \rightarrow \, e^{i
\delta (T_1 + \pi / \Omega)}$. We suppose the atom $A_1$ to interact
with mode $C_1$ for $\Omega \, t_b = \arccos{\left(
\sqrt{\frac{2}{3}} \right)}$ and afterwards $A_2$ to interact with
mode $C_1$ for $\Omega \, t_4 = \pi/2$ [cf.\ Fig.~\ref{fig:7}(a)].
This sequence together then produces the four-partite W state
\begin{widetext}
\begin{eqnarray} \label{w_d_3}
   && \ket{\Phi_4 (T_1)} = \frac{1}{\sqrt{2}}
      \left[ \frac{1}{\sqrt{2}}
             \left( \im \ket{0,\bar{1}}
              + e^{i \delta (T_1 + \frac{3\pi}{2 \Omega} + t_b)}
                    \ket{1, \bar{0}}
         \right) \ket{g_1, g_2} \right.
      % \notag \\    &&
      \left. - \frac{1}{\sqrt{2}}
               e^{i \delta (T_1 + \frac{3\pi}{2\Omega} + t_b)}
           \left( \ket{g_1, e_2} + \ket{e_1, g_2} \right)
               \ket{0, \bar{0}}
      \right]
\end{eqnarray}
\end{widetext}
for the atoms $A_1,\, A_2$ and the two cavity modes $C_1,\, C_2$.
When the atoms have left the cavity, they are projected one-by-one
in the longitudinal basis inside the detector. We are interested
only in those cavity wave packets, for which the atoms $A_1,\, A_2$
have been detected in their ground state, and hence, we shall
discard all others events right from the beginning. Using such a
state-selective procedure, we then know that the state (\ref{w_d_3})
of the atoms is reduced to the photonic Bell state
\begin{equation} \label{w_d_4}
   \ket{\Phi_5 (T_1)} = \frac{1}{\sqrt{2}}
   \left( \im \ket{0, \bar{1}} +
          e^{i \delta (T_1 + \frac{3\pi}{2 \Omega} + t_b)} \ket{1, \bar{0}}
   \right) \,
\end{equation}
to which a series of longitudinal measurements can be applied.
Notice that the duration of the first time delay $T_1$ is `stored'
in the phase of (\ref{w_d_3}) and subsequently in the phase of
(\ref{w_d_4}). Below, we show how this phase appears as a parameter
of the (measured) probability amplitude, and thus, the observed
signal provides us with `knowledge' about the coherence of the
initial state (\ref{w_d_3}) before it has been projected upon the
ground state $\ket{g_1, g_2}$. At the same time, in order to reveal
the co\-he\-rent superposition of the (entangled) cavity states
(\ref{w_d_4}), a second delay time $T_2$ is introduced in the
sequence, before the probe atom $A_p$ enters the cavity [see
Fig.~\ref{fig:7}(a)]. This free time evolution of the cavity field
state (\ref{w_d_4}) during the delay $T_2$ produces the additional
phase shift $ e^{i \delta (T_1 + 3\pi / 2 \Omega + t_b)} \,
\rightarrow \,
  e^{i \delta (T_1 + T_2 + 3\pi / 2 \Omega + t_b)}$, where the durations
of delays $T_1$ and $T_2$ are manipulated independently.

As in previous Section, while the probe atom $A_p$ crosses the
cavity, it interacts with the mode $C_1$ for $\Omega \, t_6 = \pi$
and with $C_2$ for $\Omega \, t_7 = \pi/2$, respectively. These
steps together yield the atom-cavity state
\begin{eqnarray} \label{w_d_5}
   \ket{\Phi_7 (T_1, T_2)} & = & \frac{1}{2}
   \left[ \im \left( 1 - e^{i \delta (T_1 + T_2 + \theta)} \right)
          \ket{g_p, 0, \bar{1}}
   \right.
   \nonumber \\
   & & +
   \left. \left( 1 + e^{i \delta (T_1 + T_2 + \theta)} \right)
          \ket{e_p, 1, \bar{0}} \right] ,
\end{eqnarray}
with
$\theta = (\frac{5\pi}{2} + \arccos{\left( \sqrt{\frac{2}{3}} \right)})
          /\Omega$. Moreover, the final-state probability to find the
probe atom $A_p$ in its excited state is given by
\begin{equation} \label{prob1}
   P(e_p; \, T_1, T_2) =
   \frac{1 + \cos \left( \delta (T_1 + T_2 + \theta) \right)}{2} \, ,
\end{equation}
i.e.\ by an expression containing two tunable delay times $T_1$ and
$T_2$ introduced in order to reveal the properties of the initial
four-partite W state (\ref{w_d_3}). We note once again that, while
$T_2$ helps to analyze the entanglement of the Bell state
(\ref{w_d_4}), $T_1$ is utilized to control the accuracy of
coherence transfer from the state (\ref{w_d_3}) to that of
(\ref{w_d_4}).

As before, this (temporal) sequence has to be repeated many times in
order to reconstruct the final-state pro\-ba\-bi\-li\-ty $P(e_p; \,
T_1, T_2)$ as function of $T_1$ and $T_2$. Note that the
state-selective measurements should be here again used as to collect
only those probabilities, for which the probe atom has been detected
in its exited state. If the four-partite W state from Section~II.B
was produced, this probability (distribution) should of course be
reasonably close to the predictions in Eq.~(\ref{prob1}).

\section{Summary and Outlook}

In this work, two schemes are suggested to ge\-ne\-ra\-te
four-partite entangled GHZ and W states within the framework of
cavity QED. They are based on the \mbox{resonant} interaction of (a
chain of) Rydberg atoms with a bimodal cavity that supports two
independent modes of the photon field. In addition, we show how
these schemes can be extended towards the generation of $N-$partite
GHZ and W states. To reveal the entanglement of produced states, we
also propose the (temporal) sequences of projective measurements and
time delays. Using the language of temporal sequences and quantum
circuits, a comprehensive description of all necessary
manipulations, has been achieved. Our goal is to provide a scheme
that can readily be adopted for cavity QED experiments
\cite{haroche1, haroche, haroche2} and, in particular, for a
forthcoming generation of high-finesse microwave cavities
\cite{apl90}.

Since the experimental reports \cite{prl75, pra64, prl92}, the use
of bimodal cavities has been found an important step towards the
manipulation and control of (rather) complex quantum states. A
number of proposals \cite{jmo51, pra66_1, cp16, pra67, pra68, pra76,
pla339, jpb} has been made in the literature to exploit further
capabilities of bimodal cavities. For instance, in contributions
\cite{jmo51, pra66_1, cp16, pra67} the schemes for the engineering
of various (multi-partite) entangled states between the atomic
(chain) and/or photonic qubits have been proposed. In contrast to
the present work, however, most of the previous suggestions were not
well adopted to the recent design of the cavities, and no
satisfactory attempt was made to reveal the non-classical
correlations belonging to the produced states. Another fruitful
branch of bimodal cavity applications characterizes the proposal by
Zubairy \etal \, \cite{pra68}, where a bimodal cavity is utilized to
realize a quantum phase gate in which the quantum register is
represented by the two cavity mode states. Based on this gate, the
authors suggested a scheme that enables one to implement Grover's
search algorithm by means of a bimodal cavity. We also mention the
papers \cite{pra76, pla339} where it has been demonstrated that the
coupling of both cavity modes to a common reservoir induces the
tunneling of a field state from one cavity mode to another mode of
the same cavity device, and thus, opens a way to implement the
environment assisted (short-distance) teleporting inside a bimodal
cavity. To achieve this goal, i.e. to follow the time evolution of
such quantum systems embedded into a reservoir or under the external
noise and to analyze different (entanglement or separability)
measures, a `quantum simulator' has been developed recently in our
group \cite{Radtke:06} that can be utilized for such studies in the
future.

Finally, let us recall here that all pure (genuine) four-partite
entangled states, based on qubits, can be classified into nine
classes \cite{pra65_1} by using LOCC transformations. Among these
classes, we obviously find the four-partite GHZ and W states as
discussed above. Therefore, an interesting task is to develop
schemes that enable one to generate a complete set of genuine
entangled states in the framework of cavity QED.

\begin{acknowledgments}
This work was supported by the DFG under the project No. FR 1251/13.
\end{acknowledgments}


\begin{thebibliography}{35}

%
%
%basic literature

\bibitem{nc}
 M.~A.~Nielsen, I.~L.~Chuang,
 \textit{Quantum Computation and quantum Information},
 (Cambridge University Press, Cambridge, 2000).

\bibitem{bew}
 C.~H.~Bennett and S.~J.~Wiesner,
 Phys. Rev. Lett. \textbf{69}, 2881 (1992).

\bibitem{eke}
 A.~K.~Ekert,
 Phys. Rev. Lett. \textbf{67}, 661 (1991).

\bibitem{gro}
 L.~K.~Grover,
 Phys. Rev. Lett. \textbf{79}, 325 (1997).

\bibitem{ghz}
 D.~M.~Greenberger, M.~Horne, and A.~Zeilinger, in
 \textit{Bell's Theorem, Quantum Theory, and Conceptions of the Universe},
 edited by M.~Kafatos, 73 (Kluwer, Dordrecht, 1989).

\bibitem{prl65}
 N.~D.~Mermin,
 Phys. Rev. Lett. \textbf{65}, 1838 (1990).

\bibitem{prl67}
 S.~M.~Roy, V.~Singh,
 Phys. Rev. Lett. \textbf{67}, 2761 (1991)

\bibitem{pra62}
 W.~D\"{u}r, G.~Vidal, and J.~I.~Cirac,
 Phys. Rev. A, \textbf{62}, 062314 (2000).

%
% the next block
%


\bibitem{prl_nat}
 D.~Bouwmeester, J.~W.~Pan, M.~Daniell, H.~Weinfurter, A.~Zeilinger,
 Phys. Rev. Lett. \textbf{82}, 1345 (1999); \\
 J.-W.~Pan et al.,
 Nature \textbf{403}, 515 (2000).

\bibitem{prl_jmo}
 M.~Eibl, N.~Kiesel, M.~Bourennane, C.~Kurtsiefer, H.~Weinfurter,
 Phys. Rev. Lett. \textbf{92}, 077901 (2004); \\
 M.~Kiesel et al.,
 J. of Modern Optics \textbf{50}, 1131 (2003).

\bibitem{pra61}
 R.~J.~Nelson, D.~G.~Cory, S.~Lloyd,
 Phys. Rev. A \textbf{61}, 022106 (2000).

\bibitem{pra66}
 G.~Teklemariam et al.,
 Phys. Rev. A \textbf{66}, 012309 (2002).

\bibitem{sc288}
 A.~Rauschenbeutel et al.,
 Science \textbf{288}, 2024 (2000).

\bibitem{no}
 3-W states have not been produced experimentally yet.

\bibitem{sc304}
 C.~F.~Roos et al.,
 Science \textbf{304}, 1478 (2004).

\bibitem{haroche}
 J.~M.~Raimond, M.~Brune, and S. Haroche
 {\it Rev. Mod. Phys.} \textbf{73}, 565 (2001).

\bibitem{haroche1}
 S.~Haroche and J.~M.~Raimond, in
 \textit{Cavity Quantum Electrodynamics},
 edited by P.~Bergman, 123 (Academic Press, 1994).

\bibitem{haroche2}
 S.~Haroche, J.~M.~Raimond,
 \textit{Exploring the Quantum: Atoms, Cavities, and Photons}
 (Oxford Univeristy Press, 2006)

\bibitem{apl90}
 S.~Kuhr et al.,
 Appl. Phys. Lett. \textbf{90}, 164101 (2007).

\bibitem{jc}
 E.~T.~Jaynes and F.~W.~Cummings,
 Proc. IEEE \textbf{51}, 89 (1963).

\bibitem{prl86}
 J.-W.~Pan, M.~Daniell, S.~Gasparoni, G.~Weihs, A.~Zeilinger,
 Phys. Rev. Lett. \textbf{86}, 4435 (2001).

\bibitem{nat404}
 C.~A.~Sackett et al.,
 Nature \textbf{404}, 256 (2000).

\bibitem{prl90}
 M.~Eibl et al.,
 Phys. Rev. Lett. \textbf{90}, 200403 (2003).

\bibitem{prl79}
 E.~Hagley, et al.,
 Phys. Rev. Lett. \textbf{79}, 1 (1997).

\bibitem{pra64}
 A.~Rauschenbeutel et al.,
 Phys. Rev. A \textbf{64}, 050301(R) (2001).

\bibitem{prl75}
 Q.~A.~Turchette, C.~J.~Hood, W.~Lange, H.~Mabuchi, H.~J.~Kimble,
 Phys. Rev. Lett. \textbf{75}, 4710 (1995).

\bibitem{prl92}
 L.~M.~Duan and H.~J.~Kimble,
 Phys. Rev. Lett. \textbf{92}, 127902 (2004).


%
% next block of refs
%


\bibitem{pr47}
 A.~Einstein, B.~Podolsky, and N.~Rosen,
 Phys. Rev. \textbf{47}, 777 (1935).

\bibitem{jmo51}
 A.~Biswas, G.~S.~Agarwal,
 J. of Modern Optics \textbf{51}, 1627 (2004).

\bibitem{pra66_1}
 M.~Ikram, F.~Saif,
 Phys. Rev. A \textbf{66}, 014304 (2002).

\bibitem{cp16}
 Y.~Zhen-Biao and S.~Wan-Jun,
 Chinese Phys. \textbf{16}, 435 (2007).

\bibitem{pra67}
 C.~Wildfeuer and D.~H.~Schiller,
 Phys. Rev. A \textbf{67}, 053801 (2003).

\bibitem{pra68}
 M.~S.~Zubairy, M.~Kim, and M.~.O.~Scully,
 Phys. Rev. A \textbf{68}, 033820 (2003).

\bibitem{pra76}
 I.~P.~de Queiros, S.~Souza, W.~B.~Cardoso, and N.~G.~de Almeida
 Phys. Rev. A \textbf{76}, 034101 (2007).

\bibitem{pla339}
 A.~R.~Bosco de Magalhaes and M.~C.~Nemes
 Phys. Lett. A \textbf{339}, 294 (2005).

\bibitem{jpb}
 D.~Gonta and S.~Fritzsche,
 J. Phys. B: At. Mol. Opt. Phys. \textbf{41}, 095503 (2008).

\bibitem{notice}
 In order to be compatible with the Bloch-sphere
 notation of Section III, the fourth qubit should read
 $\keta{\ua_4} = \ket{e_4}$ and $\keta{\da_4} = \ket{g_4}$.
 In Section II, however, we can safely use the `reversed'
 notation from Eqs. (\ref{ghz_basis}) in order to represent
 the 4-GHZ state (\ref{ghz_p_end}) in the form as it was
 previously announced in Eq. (\ref{4-ghz}).


%
% next block of refs
%

\bibitem{Radtke:06}
 T.~Radtke and S.~Fritzsche,
 Comput. Phys. Commun. \textbf{175}, 145 (2006);
 \textit{ibid.}        \textbf{176}, 617 (2007).

\bibitem{pra65_1}
 F.~Verstraete, J.~Dehaene, B.~DeMoor, H.~Verschelde,
 Phys. Rev. A \textbf{65}, 052112 (2002).


\end{thebibliography}
\end{document}